\documentclass[prd,aps,twocolumn,nofootinbib]{revtex4}
\usepackage{graphicx,url}
\usepackage{amsmath,amssymb}
\usepackage{array}

\def\be{\begin{equation}}
\def\ee{\end{equation}}
\def\ba{\begin{align}}
\def\ea{\end{align}}

\def\lsim{\raise0.3ex\hbox{$\;<$\kern-0.75em\raise-1.1ex\hbox{$\sim\;$}}}
\def\gsim{\raise0.3ex\hbox{$\;>$\kern-0.75em\raise-1.1ex\hbox{$\sim\;$}}}

\def\theta{\vartheta}

\def\R{{\cal R}}

\newcommand{\mnras}{{Mon.\ Not.\ Roy.\ Astron.\ Soc. }}

\renewcommand{\vec}[1]{\boldsymbol{#1}}

\begin{document}

\title{High-energy neutrinos from cosmic ray interactions in the Local Bubble}

\author{M.~Bouyahiaoui$^{1}$}
\author{M.~Kachelrie\ss$^{2}$}
\author{D.~V.~Semikoz$^{1}$}

\affiliation{$^{1}$APC, Universit\'e Paris Diderot, CNRS/IN2P3, CEA/IRFU,
Observatoire de Paris, Sorbonne Paris Cit\'e, 119 75205 Paris, France}
\affiliation{$^{2}$Institutt for fysikk, NTNU, Trondheim, Norway}

\begin{abstract}
A surprisingly large flux of extraterrestrial high-energy neutrinos was
discovered by the IceCube experiment. While the flux of muon neutrinos
with energies $E>100$\,TeV is consistent with the extragalactic gamma-ray
background (EBL) determined by Fermi-LAT, the softer component of the
cascade neutrino flux
at $E<100$\,TeV is larger than expected. Moreover, a gamma-ray
excess at high Galactic latitudes at energies $E>300$\,GeV was found in
the data of Fermi-LAT. The gamma-ray excess at TeV energies and the neutrino
excess at $E<100$\,TeV may have a common Galactic origin. In this work, we
study the possibility that both excesses are caused by interactions of cosmic
rays (CRs) with energies up to PeV in the wall of the Local Bubble. Source of
these CRs may be a recent nearby  source like Vela.  We show that such a
scenario can explain
the observed CR flux around the knee, while CR interactions in the
bubble wall can generate a substantial fraction of the observed astrophysical
high-energy neutrino flux below  $\sim {\rm few} \times 100$\,TeV. 
\end{abstract}

\maketitle

\section{Introduction}

High-energy astrophysical neutrinos are a unique probe to understand the
non-thermal universe~\cite{Gaisser:1994yf}. They are produced together with
photons in interactions of cosmic rays (CR) on matter and background photons
close to their sources and during propagation. Since neutrinos travel
undisturbed, being neither absorbed as high-energy photons nor deflected in
magnetic fields as charged particles, they are an ideal tracer of CR
sources.

The IceCube collaboration employs two main experimental channels to detect such
neutrinos. Using the tracks of muons one can measure the neutrino arrival
direction rather precisely, while its energy can be only estimated within a
factor of a few~\cite{Aartsen:2016xlq}. To avoid the atmospheric neutrino
background, the energy spectrum of astrophysical neutrinos in this channel is
measured above 200\,TeV. A fit with a power law to the flux of muon neutrinos
from 9.5~years of observations resulted in~\cite{Aartsen:2020aqd} 
\begin{equation}
E^2 F(E) = (4.32 \pm 0.9) \times  10^{-8}\left( \frac{E}{100\, \rm TeV} \right)^{{-0.3 \pm 0.1}}  \frac{\rm  GeV}{\rm cm^2 \, s \, sr}.
\label{flux_muon}
\end{equation}
The slope  is consistent with a power law $1/E^\alpha$ with index
$\alpha=2.0$--2.2 which is predicted in many models of extragalactic neutrino
sources~\cite{Stecker:1991vm,Mannheim:1995mm,Waxman:1997ti,Loeb:2006tw}.
Moreover, the magnitude of the flux agrees with
the one expected from the  extragalactic gamma-ray background (EBL), which
should contain a comparable energy~\cite{Berezinsky:2010xa,Murase:2013rfa}.

In a second channel using cascade events inside the IceCube detector one
can measure electron neutrinos interacting via the charged current and
additionally all neutrino flavours interacting via the neutral current.
The energy spectrum of astrophysical neutrinos derived
in this channel per neutrino flavor is~\cite{Aartsen:2020aqd} 
\begin{equation}
E^2 F(E) = (4.92 \pm 1.1)\times 10^{-8}\left( \frac{E}{100 \,\rm TeV} \right)^{-0.53 \pm 0.1}  \frac{\rm  GeV}{\rm cm^2 \, s \, sr} .
\label{flux_cascade}
\end{equation}
Such a
steep spectrum challenges an extragalactic origin of this component, since
the accompanying photons would overshoot the bounds on the diffuse background
of extragalactic gamma-rays~\cite{Berezinsky:2010xa,Murase:2013rfa}.

In Ref.~\cite{Neronov:2014uma}, it was noticed that the all-sky  gamma-ray 
flux measured by Fermi-LAT is consistent with the soft neutrino spectrum in
Eq.~(\ref{flux_cascade}). While the observed gamma-ray flux is dominated by
the contribution of the Galactic plane, a corresponding neutrino contribution
is tightly constraint by data from both IceCube and
ANTARES~\cite{Albert:2017oba}. Therefore a Galactic neutrino contribution
should be rather isotropic, with a significant fraction of the flux coming
from outside of the Galactic plane.

In Ref.~\cite{Neronov:2015osa}, evidence for a Galactic contribution in the
HESE neutrino data was found based on both the signal in the Galactic plane
and at high Galactic latitudes.  Two-component models with a Galactic and 
an extragalactic contribution were suggested in Refs.~\cite{Neronov:2016bnp}
and \cite{Palladino:2017qda} to explain the data in both the muon and cascade
channels. A non-zero Galactic contribution was obtained
also more recently in a multi-component fit performed in 
Ref.~\cite{Palladino:2018evm}. Finally, a gamma-ray excess at high Galactic
latitudes in the Fermi LAT data with energies $E>300$\,GeV was reported in
Ref.~\cite{Neronov:2018ibl,Neronov:2019ncc}. An extension of the
Fermi gamma-ray excess to energies of  $\gg 10$\,TeV would indicate its
Galactic origin, since at these energies photons are strongly attenuated by
pair-production on cosmic microwave background and  IR photons.

One possible explanation for a close to isotropic Galactic neutrino flux is
that the neutrinos originate from a large Galactic halo, formed either by
CRs~\cite{Taylor:2014hya,Blasi:2019obb} (but see Ref.~\cite{Kalashev:2016euk})
or heavy dark matter
particles~\cite{Berezinsky:1997hy,Feldstein:2013kka,Esmaili:2013gha}.
Another proposed Galactic neutrino source are the Fermi bubbles,
 if these gamma-rays have an hadronic origin as suggested, e.g., in Refs.~\cite{Crocker:2010dg,Lunardini:2011br}. 
Finally, there is the possibility that the main contribution to the Galactic
neutrino
flux is rather local, produced by CRs interactions in the walls of the Local
Bubble~\cite{Andersen:2017yyg}.

A smoking gun of all Galactic models explaining the IceCube data below
100\,TeV is the associated, nearly isotropic flux of 
 photon with energies 10--100\,TeV.
Such a photon flux can be measured by dedicated experiments, which are
able to suppress the CR background by a factor $f_{\rm CR}\lsim 3 \times 10^{-5}$.
The required suppression factor corresponds to the ratio between the neutrino
and CR flux in this energy range, and should be similar or better than the one achieved in
the KASCADE measurements. Examples for such experiments are the existing
CARPET-3 facility~\cite{Dzhappuev:2018bnl} and the next generation experiment
LHAASO~\cite{DiSciascio:2016rgi}, which is currently in the construction stage.

Once a diffuse gamma-ray flux above 10\,TeV will be detected, its angular
distribution can help to distinguish between the various proposed models.
Signature of heavy dark matter is an excess towards the direction of the
Galactic center. The gamma-ray flux from CRs in an extended Galactic halo
should be nearly isotropic, apart from an energy dependent suppression towards
the Galactic center due to pair production. Finally, the gamma-ray flux
predicted in models of local CR sources should follow the distribution of gas
in the Local Bubble and its wall. In addition, large scale anisotropies in
the TeV--PeV energy ange have been observed in the, e.g., IceTop and IceCube
data~\cite{Desiati:2013lea}. These anisotropies add further constraints on
the origin of cosmics rays producing neutrinos and gamma rays.

In this work, we study the latter possibility assuming that a young CR source,
as e.g.\ Vela, is located outside the Local Bubble and contributes
significantly to the locally observed CR flux in the energy region of the
CR knee  ~\cite{Bouyahiaoui:2018lew}. We show that CR interactions in the
wall of the Local Bubble can lead to a bump in the neutrino flux around
10\,TeV, consistent with the soft spectrum from Eq.~(\ref{flux_cascade}),
while the accompanying photon flux is below the limits set by Kascade-Grande. 

This work is structured as follows: We describe first in Sec.~\ref{bubble}
our model for the local magnetic field and the geometry of the Local Bubble.
Then we discuss in Sec.~\ref{source} the calculation of the CR flux in
the wall and the interior of the bubble, as well as their dependence on the
parameters describing the bubble wall. Having fixed these parameters, we
present in
Sec.~\ref{neutrino} the resulting CR, neutrino and photon fluxes and compare
them to the observations. Finally, we conclude in Sec.~\ref{concl}.

\section{Local Bubble and the geometry of the local magnetic field}
\label{bubble}
  
In Ref.~\cite{Bouyahiaoui:2018lew}, we proposed a model in which the CR flux
around the knee is dominated by the contribution of the Vela supernova remnant.
In that work, which we will denote in following as case U, we
used a toy model for the magnetic field structure in and around the Local
Bubble (LB) inspired by Ref.~\cite{Andersen:2017yyg}. In particular, we
assumed  cylindrical symmetry and approximated the field outside the bubble
with radius $R=100$\,pc as uniform. Additionally, we employ now as
case JF a more realistic description of the magnetic field
outside the bubble using the Jansson-Farrar (JF) model for the Galactic
magnetic field~\cite{Jansson:2012rt}, with the strength of the
turbulent field reduced to obey B/C measurements as discussed in
Refs.~\cite{Giacinti:2014xya,Giacinti:2015hva}.
We assume that the strength of the regular magnetic field inside the bubble
depends only on the radius $r$ from the center of the bubble and the height
$z$ above the plane, setting $B_{\rm in}=0.1 \mu$G inside
the bubble and $B_{\rm sh}=(8 - 12)\mu$G in the wall for $z=0$.
Then we apply an exponential damping of the magnetic field inside the
bubble as function of the distance $z$ to the Galactic plane with height
scale $z_{\rm b}= 100 $\,pc, since the Local Buble extends to
  high latitudes, $|b|\geq 35^{\circ}$~\cite{Welsh:2009sg}.
For the geometry of the
bubble field, we assume inside the bubble and
the wall a clockwise oriented magnetic field for $y > 0$ and an anticlockwise
one for $y<0$. 

The transitions between different magnetic field regimes are interpolated by
logistic functions $T(r)$. The width of the two transitions is parametrised 
by $w_i$ with $i=\{1,2\}$, while $w$ denotes the extension of the wall. We
will discuss the dependence of our results on the chosen value of these
parameters in Sec.~\ref{dependence}. As our default parameters, we
use $w=2$\,pc, $w_1=1$\,pc, and $w_2=0.1$\,pc, while we set as strength of
the regular magnetic field in the wall $B_{\rm sh}=12\mu$G, $B_{\rm in}=0.1\,\mu$G
inside and $B_{\rm out}=1\,\mu$G outside the bubble, respectively. Then
the transition functions are given by 
\begin{align}
T_1 & = \left[1+ \exp\left(-\frac{r-R+w/2}{w_1}\right) \right]^{-1}, \\
T_2 & = \left[1+ \exp\left(-\frac{r-R-w/2}{w_2}\right) \right]^{-1},
\end{align}
where we identify the center of the LB with the origin of our coordinate
system. We set the $x$,$y$ and $z$ components of the magnetic field for
$r<R$ to 
\begin{align}
\begin{split}
B_{\rm x}  & =  s \left[B_{\rm in} (1-T_1)+B_{\rm sh}T_1\right]\sin(\theta)  \exp(-z^2/z^2_{\rm b} )
\\
& + B_{\rm JF x}  (1-\exp(-z^2/z^2_{\rm b}) ),
\end{split}
\\
\begin{split}
 B_{\rm y}  & =  -s \left[B_{\rm in} (1-T_1)+B_{\rm sh}T_1\right]\cos(\theta)  \exp(-z^2/z^2_{\rm b} )
\\
& + B_{\rm JF y}  (1-\exp(-z^2/z^2_{\rm b}) ),
\end{split}
\\
\begin{split}
 B_{\rm z}  & =  B_{\rm JF z}  .
 \end{split}
\end{align}
Similarly, the field is given for $r >R$ by
\begin{align}
\begin{split}
B_{\rm x}  & =   \left[s  B_{\rm sh} (1-T_2) \sin(\theta) +B_{\rm JF x}T_2\right]  \exp(-z^2/z^2_{\rm b} )
\\
& + B_{\rm JF x}  (1-\exp(-z^2/z^2_{\rm b}) ),
\end{split}
\\
\begin{split}
 B_{\rm y}  & =   \left[-s  B_{\rm sh} (1-T_2) \cos(\theta) +B_{\rm JF y}T_2\right] \exp(-z^2/z^2_{\rm b} )
\\
& + B_{\rm JF y}  (1-\exp(-z^2/z^2_{\rm b}) ),
\end{split}
\\
\begin{split}
 B_{\rm z}  & =  B_{\rm JF z}  ,
 \end{split}
\end{align}
%
%
%
with $s=y/|y|$ and $B_{\rm JF}$ are the corresponding components of the
field in the JF model.

The turbulent magnetic field modes are distributed between
$L_{\rm min}=1$\,AU and $L_{ \rm max} = 25$\,pc according to an isotropic
Kolmogorov power spectrum. We construct the turbulent magnetic field using
nested grids as described in Ref.~\cite{Giacinti:2011ww}. In the actual
simulations, only field modes above $L_{\rm min}^\prime = 0.01$\,pc  were
included. Inside the wall, we omit all Fourier modes with $L>L_{\max}/100$,
such that the largest modes still have few oscillations within the thickness
of the wall. To maintain the same strength $B_{\rm rms}$ of the turbulent field,
we enhance the power in the modes with smaller wave-lengths.
The strength of the turbulent field is set for
$(R-w/2-3w_1) \le r \le (R+w/2+3w_2)$ to  
$B_{\rm turb}= B_{\rm reg}/2$, and for $r <(R-w/2-3w_1)$ to
$B_{\rm turb}=5B_{\rm in}$. 
Hence CR propagate outside the bubble 
anisotropically, but inside nearly isotropically. This choice is motivated
by the notion that the supernova explosion which created the LB
expelled the regular field into the wall. At the same time, the injected
turbulence lead to an increase of the turbulent component of
the magnetic field in the bubble.

The Sun is assumed to be at the central part of the LB, while Vela is
situated at the coordinates $r_{\rm Vela}= 0.29$\,kpc,
$l_{\rm Vela}= -3.4^\circ$ and $b_{\rm Vela} = 263.9^\circ$. This implies
that the Sun and Vela are connected by a regular magnetic field line in
the original JF model. In the case of anisotropic diffusion, the CR density
is strongly enhanced at small perpendicular distances to the field line
through the source~\cite{Giacinti:2017dgt}. Therefore the CR flux from
Vela typically overshots the observed flux in the knee region in such models. 
As we will show in the next section, the strong field in the bubble wall
acts as a CR shield, reducing the locally observed CR flux inside the
bubble.

\section{Model of a local neutrino source}
\label{source}

In this section, we discuss the chosen CR injection spectrum and
the calculation of the resulting CR fluxes, as well as the dependence
of the CR flux on the parameters of the bubble.

\subsection{Motivation for the choice of the source}

\begin{figure}[h!]
  \hspace*{-0.7cm}
  \includegraphics[width=0.55\textwidth]{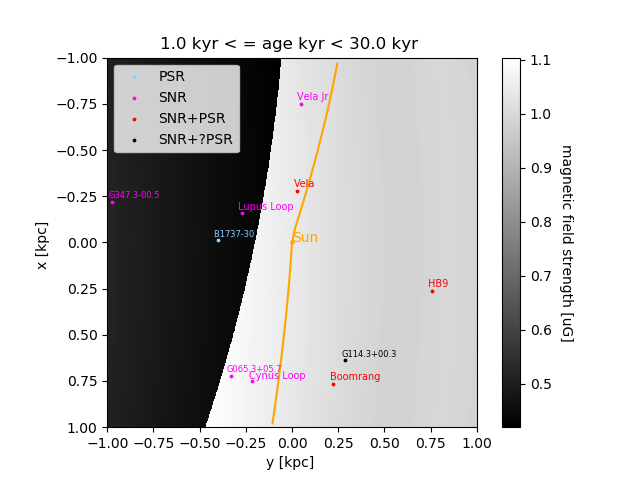}
  \caption{Location of potential CR sources with age between 1--30\,kyr.
    The Galactic center is at $(x,y)=(0,-8.5\,{\rm kpc})$. The yellow line
    shows the magnetic
    field line passing through the Sun projected on the Galactic plane. 
   \label{fig:fig_sources}}
\end{figure}

Cosmic rays propagate preferentially along the regular magnetic field lines.
Therefore, the number of CR sources contributing to the locally observed CR
flux is reduced relative to the case of isotropic
diffusion~\cite{Giacinti:2017dgt}. In particular, the contribution of a source
to the local flux depends strongly on its perpendicular distance to the
magnetic field line through the Sun. In Fig.~\ref{fig:fig_sources}, we show
the position of potential CR sources which are younger than 30\,kyr.
 As potential sources
we distinguish pulsars (light blue points), SNRs (magenta points), SNRs associated
with pulsar (red points), and SNRs possibly associated with pulsars (black
points). Additionally, we show the magnetic field line passing through the Sun
in the JF model as a yellow line. We note that the kink of the field line
near the position of the Sun is a projection effect, caused by the $z$
component of the field. Only four sources are close to the magnetic field line
through the Sun; these sources are in the position to give a dominating
contribution to the local CR flux.

\begin{table}[h!]
\begin{center}
\begin{tabular}{c|c|c|c|c|c|c}
 Source & Name  & $\tau$/kyr &$d$  & $d_B$ & $\Delta_\|$ & $\Delta_\perp$\\ 
  \hline
   G065.3+05.7 & -- & 20 & 0.8  &  0.27 & 0.96 & 0.036 \\ 
    \hline
  G074.0-08.5 & Cygnus Loop & 15 & 0.78  &  0.3 & 0.83 & 0.032 \\ 
    \hline
    G106.3+02.7 & Boomerang  & 10 & 0.8   &  0.32 & 0.68  & 0.026 \\ 
    \hline
    G114.3+00.3 & --  & 7.7   & 0.7 &   0.36 & 0.6 & 0.023 \\ 
  \hline
   G160.9+02.6 & HB9 & 5.5  & 0.8   &   0.77 & 0.5 & 0.019 \\ 
  \hline
   G263.9-03.3 & Vela & 11   & 0.29 &   0.06 & 0.71 &  0.027 \\ 
  \hline
   G266.2-01.2 & Vela Jr  & 3.8 & 0.75   &   0.18 & 0.42 & 0.016 \\ 
  \hline
   G330.0+15.0 & Lupus Loop& 23   & 0.33  &   0.32 & 1.03  & 0.039 \\ 
  \hline
   G347.3-00.5 & -- & 1.6   & 1  &   0.8 & 0.27  & 0.01 \\ 
  \hline
   B1737-30 & -- & 20.6  & 0.4  &   0.4 & 0.98 & 0.037  \\ 
\end{tabular}
\end{center}
\caption{Properties of the sources in Fig.~\ref{fig:fig_sources}:
  $d$ and $d_B$ denote the distance to the Sun and to the magnetic field line
  passing through the Sun, respectively, while $\Delta_i$ is the typical
  distance CRs with energy $E=3$\,PeV diffuse parallel and perpendicular to
  the magnetic field line. All distances are in kpc.\label{table:tab_sources}}
\end{table}

Table~\ref{table:tab_sources} summarises the available information\footnote{Data on the pulsars are from \url{https://www.atnf.csiro.au/people/pulsar/psrcat/} and on SNRs from \url{http://snrcat.physics.umanitoba.ca/SNRtable.php}.}
on the CR sources shown in Fig.~\ref{fig:fig_sources}.
The last column shows the distance $d_i=\sqrt{2D_i\tau}$ beyond which the 
CR flux from a CR source with age $\tau$ is exponentially suppressed. Since
the local magnetic field line is approximately aligned with the $x$ axis,
we can set $\Delta_x\simeq\Delta_\|\simeq\sqrt{2D_\|\tau}$ and
$\Delta_{y,z}\simeq\Delta_{\perp}\simeq\sqrt{2D_\perp \tau}$,
with $D_\|$ and $D_\perp$ as the parallel and perpendicular diffusion
coefficients, respectively. These diffusion coefficient were computed
numerically for the position of Vela  and the CR energy $E=3$\,PeV.
The typical distance $\Delta_i$ CRs with such an energy diffuse during the
time $\tau$ should be
compared to the perpendicular distance $d_B$ of the source to the magnetic
field line passing through the Sun. Depending on the ratio of these two
quantities, we can distinguish two cases: 
\begin{itemize}
\item $\Delta_\perp \gg d_B$: The contribution of this source to the local CR
  flux is exponentially suppressed.
\item $\Delta_\perp \lsim d_B$: The source contributes to the local CR flux.
\end{itemize}
From Table~\ref{table:tab_sources} it is clear that the contribution from
Vela is the dominating one.

Repeating the same analysis for older sources is less conclusive. First,
one usually does not observe the shell but only the pulsar for sources
which are few Myr old. Since pulsar velocities are typically high,
it is difficult to reconstruct the actual position of the supernova
explosion. Therefore we assume here motivated by the observations of
Fe-60~\cite{Knie:1999zz,Benitez:2002jt,Fitoussi:2007ef,2016Natur.532...69W},
that only one or two additional CR sources with an age 2--3\,Myr
contribute to the local CR flux in the energy range from $\sim 10$\,TeV
up to the knee.

\subsection{Injection spectrum and calculation of the flux}             

We use as CR injection spectrum for Vela a broken power law in rigidity
$\R=E/(Ze)$ with a break at $\R_{\rm br}=2\times 10^{15}$\,V and an exponential
cut-off at $\R_{\rm max}=8\times 10^{15}$\,V for case U, and
$\R_{\rm br}=3 \times 10^{15}$\,V and  $\R_{\rm max}=8 \times 10^{15}$\,V for
case JF, respectively,
\be
\frac{dN}{d\R} \propto \left\{
    \begin{array}{ll}
        \R^{-2.2} , & \mbox{\quad if \quad} \R < \R_{\rm br} \\
        \R^{-3.1} \exp(-\R/\R_{\rm max})), & \mbox{\quad if \quad} \R \ge \R_{\rm br}.
    \end{array}
    \right.
\ee
The steepening of the injection spectrum  by $\Delta\beta=0.9$ is
motivated, e.g., by the analysis of Ref.~\cite{Drury:2003fd}: Including
strong field amplification as suggested by Bell and
Lucek~\cite{2001MNRAS.321..433B,2004MNRAS.353..550B}
into a toy acceleration model, these authors found a break in the energy
spectrum of accelerated protons. For typical values of the SNR parameters,
this break is located close to the knee region. The strength $\Delta\beta$
of this steepening depends among others on the injection history, and in
a test particle ansatz $\Delta\beta=0.9$ was found.

The numerical values of the break and the cut-off as well as the relative
normalisation of the different groups of nuclear elements were chosen
such to reproduce best the measured CR composition. The overall normalisation
of the CR flux observed on Earth is strongly influenced by the Local Bubble,
as we will discuss in the next subsection.

In order to compute the flux, we injected 30.000~protons per energy
at the position of Vela and propagated them for 12.000\,yr. We calculated the
CR density $n(E)$ in the three regions of interest averaging the CR densities
between 8 to 12\,kyr: around the source, in the bubble wall, and inside
the bubble. The CR flux $F(E)=c/(4\pi)n(E)$ was then computed from the CR
densities in the considered volumes.
For energies below 100\,TeV we deduced the flux inside the bubble from
the flux calculated at earlier times and higher energies using the scaling
relation  
\begin{equation}
(E_{\rm low}/E_{\rm high})^{1/3} \approx t_{\rm early}/t_{\rm now} .
\label{eq_low_energy}
\end{equation}
This relation was confirmed in the numerical simulations presented in the
supplementary material of Ref.~\cite{Kachelriess:2015oua}.

\subsection{Parameter dependence of the fluxes}            
\label{dependence}

The main parameters of our model for the magnetic field in the LB are the
magnetic field strength $B_{\rm sh}$ in the wall, the wall extension $w$ and the
widths $w_{1/2}$ of the transition regions between the magnetic field in
the wall and the outside. Varying these parameters, we study how
the flux in the wall and inside the LB changes. 

To do so, we run a set of simulations modifying each time only one parameter.
To make these simulations less computing time expansive, we consider a
miniature model with a smaller bubble, $R=50$\,pc, and a reduced distance to
the source, $d=100$\,pc. Outside the bubble, we use instead of the JF model
a uniform magnetic field  directed along the $x$ axis with strength
$B_{\rm out} = 3\mu$G. Moreover, we compute the CR flux at an earlier time,
$T=7$\,kyr. Therefore, the fluxes obtained should not be compared to
experimental data, but serve to illustrate how the flux in the wall and
inside the bubble depends on the various parameters. If not otherwise
specified, we choose the widths as $w=3$\,pc,  $w_1=0.1$\,pc and
$w_2=0.01$\,pc, while we set the magnetic field strength in the wall
to $B_{\rm sh}=10\mu$G.

\subsubsection{Wall thickness}

\begin{figure}[h!]
   \begin{minipage}[b]{1.\linewidth}
      \centering \includegraphics[width=1.\linewidth]{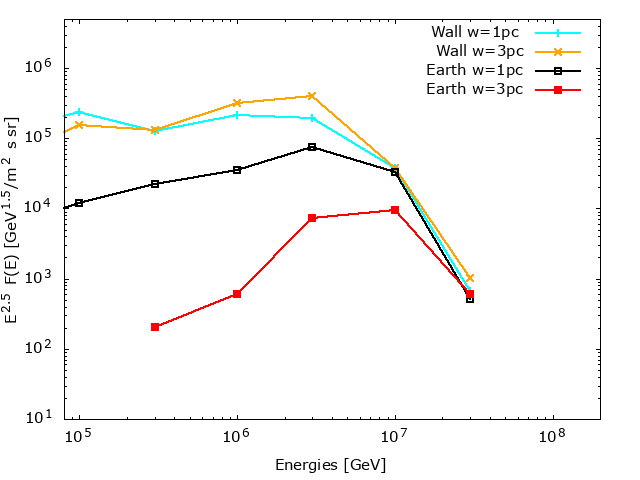}
      \caption{The proton flux computed for two different values of the
        wall thickness: $w=1$\,pc,  cyan line in the wall, and black line inside the bubble, and $w=3$\,pc,  orange line in the wall, and red line inside the bubble.
      \label{fig:w}}
   \end{minipage}\hfill
\end{figure}

Figure~\ref{fig:w} shows the flux inside the bubble and in the wall  for two different values of the wall thickness, $w=1$\,pc and
$w=3$\,pc,
respectively. While the flux of protons in the wall is practically independent
from the wall thickness $w$, the fraction of protons traversing the wall and
thus entering the bubble depends strongly on it. In contrast,
the flux in the wall practically does not change varying $w$, since only 
a small fraction of protons quits the wall and enters the bubble.

\subsubsection{Wall magnetic field amplitude}

\begin{figure}[h!]
   \begin{minipage}[b]{1.\linewidth}
      \centering \includegraphics[width=1.\linewidth]{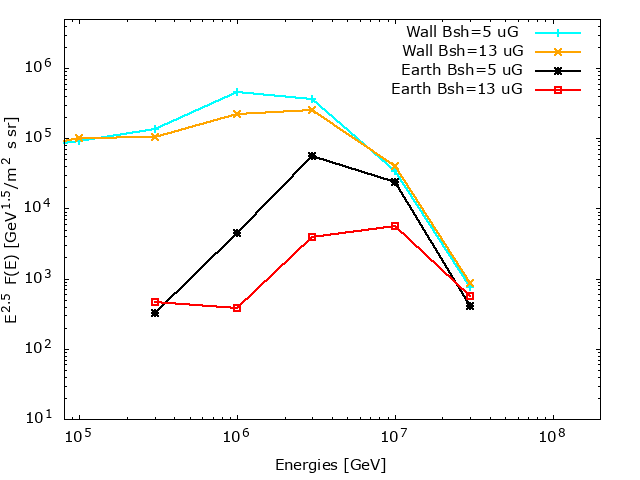}
      \caption{The proton flux computed for two different wall magnetic field $B_{\rm sh}=5$\,$\mu G$,  cyan line in the wall, and black line inside the bubble, and $B_{\rm sh}=13$\,$\mu G$,  orange line in the wall, and red line inside the bubble.
      \label{fig:uG}}
   \end{minipage}\hfill
\end{figure}

A similar behavior is found for the dependence of the fluxes on the
amplitude of the magnetic field in the wall: A stronger magnetic field
in the wall leads to a smaller  fraction of protons entering the bubble,
as they diffuse slower inside the wall. This behavior is shown in
Fig.~\ref{fig:uG},  where we plot the fluxes for two field strengths,
$B_{\rm sh}=5\,\mu$G and $B_{\rm sh}=13\,\mu$G, in the shell.
The flux inside the bubble is determined by the fraction of particles exiting
the wall. It is a function of the wall thickness, the amplitude of the magnetic
field in the wall, and the energy of the particle. For a given strength of
the magnetic field one should compare the Larmor radius $R_L$ of the particle
and the wall thickness: For $R_L\gg w$, particles cross the wall easily without
scattering, while for $R_L\ll w$ (and short enough propagation times) they
are trapped in the wall.

\subsubsection{Transition widths} 

\begin{figure}[h!]
   \begin{minipage}[b]{1.\linewidth}
      \centering \includegraphics[width=1.\linewidth]{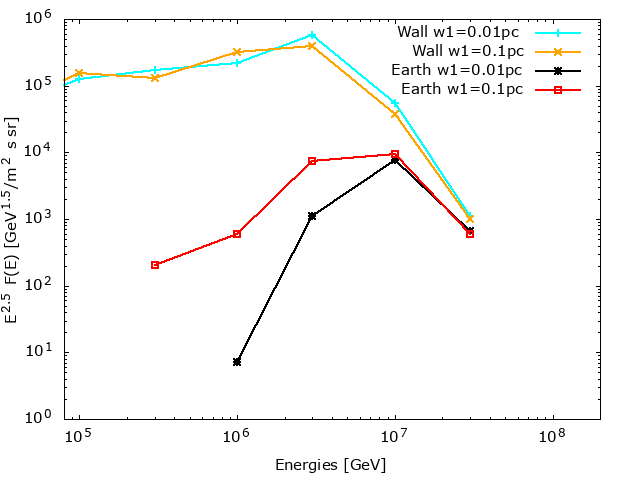}
      \caption{The proton flux computed for two different transition widths: $w_1=0.01$\,pc, cyan line in the wall, and black line inside the bubble, and $w_1=0.1$\,pc  orange line in the wall, and red line inside the bubble. Both cases with $w_2=0.01$\,pc.
      \label{fig:w1}}
   \end{minipage}\hfill
\end{figure}
\begin{figure}[h!]
   \begin{minipage}[b]{1.\linewidth}
      \centering \includegraphics[width=1.\linewidth]{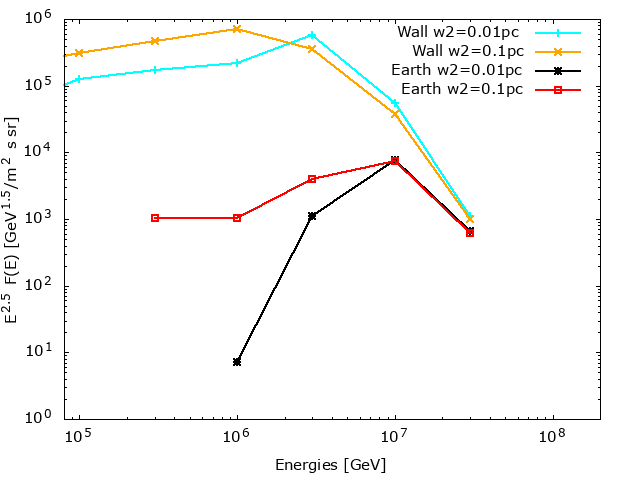}
      \caption{The proton flux computed for two different transition width $w_2=0.01$\,pc, cyan line in the wall, and black line inside the bubble, and $w_2=0.1$\,pc  orange line in the wall, and red line inside the bubble. Both cases with $w_1=0.01$\,pc.
      \label{fig:w2}}
   \end{minipage}\hfill
\end{figure}

We have seen that the CR flux in the wall  depends neither on
the field strength nor the extension of the wall. In contrast, the width
$w_2$ of the transition region between the outside and the wall influences
the CR flux in the wall: When the transition is wider, the variation of the
magnetic field strength is smaller and less protons are reflected.
In the Fig.~\ref{fig:w2}, one can see that  the flux in the wall increases
by a factor three for a transition width $w_2$ ten times larger. This implies
also a higher flux inside the bubble. The same phenomena happen varying
the second width $w_1$ between the bubble wall and the inside, as shown
in Fig.~\ref{fig:w1}: Increasing $w_1$ increases the flux inside the bubble,
because less particles are reflected.

\section{CR and secondary fluxes from Vela}            

After having discussed the parameter dependence of the CR flux, we present
next the CR and secondary fluxes from Vela for the specific geometry and
the magnetic field of the LB described in Sec.~\ref{bubble}.

\subsection{CR fluxes}            

\begin{figure}
\begin{center}
\includegraphics[width=.45\textwidth]{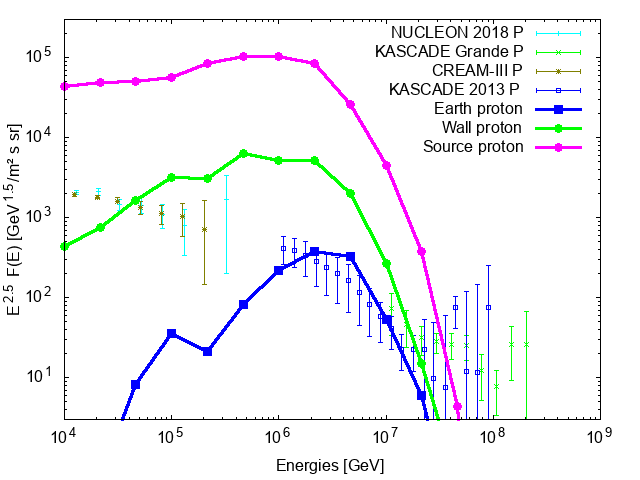}
\includegraphics[width=.45\textwidth]{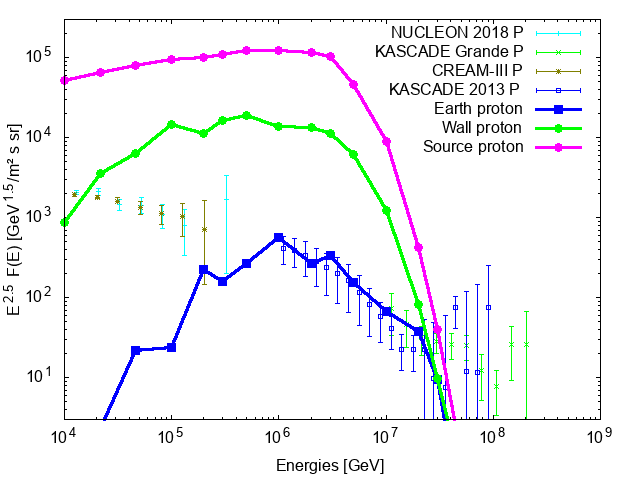}
\end{center}
\caption{Contribution of Vela to the proton flux on Earth in the model of
  Ref.~\cite{Bouyahiaoui:2018lew}. The proton fluxes at the source, in the
  wall of the Local Bubble wall and near the Earth are shown with magenta,
  green and blue lines, and compared to experimental data from
  NUCLEON~\cite{Gorbunov:2018stf}, CREAM--3~\cite{Yoon:2017qjx} as well
  as from KASCADE and KASCADE-Grande~\cite{Apel:2013uni}.}
\label{fig:protons}
\end{figure}

In Fig.~\ref{fig:protons}, we show the normalised proton flux in
the bubble wall, inside the bubble and around the source, for the
cases U (top panel) and JF (bottom panel). Our results are compared at
low energies to the data of direct  cosmic ray measurements, 
NUCLEON~\cite{Gorbunov:2018stf} and and CREAM--3~\cite{Yoon:2017qjx},
while we show at higher energies indirect measurements from 
KASCADE and KASCADE-Grande~\cite{Apel:2013uni}. At high energies,
$E\gsim 10^{16}$\,eV, the bubble wall is transparent, since
the Larmor radius ($R_L\sim 100$\,pc) of such protons is large compared to
the thickness of the bubble wall. For energies below 1\,PeV, particles start
to be  trapped in the wall and the flux inside the bubble is increasingly
suppressed. While the general behaviour in both cases is similar, the
proton flux at Earth is higher in the case JF. This difference can be
explained by the larger transition width $w_1$ we use in case JF,
$w_1 = 1$\,pc, compared to $w_1=0.1$\,pc in case U.

\begin{figure}
\begin{center}
\includegraphics[width=.45\textwidth]{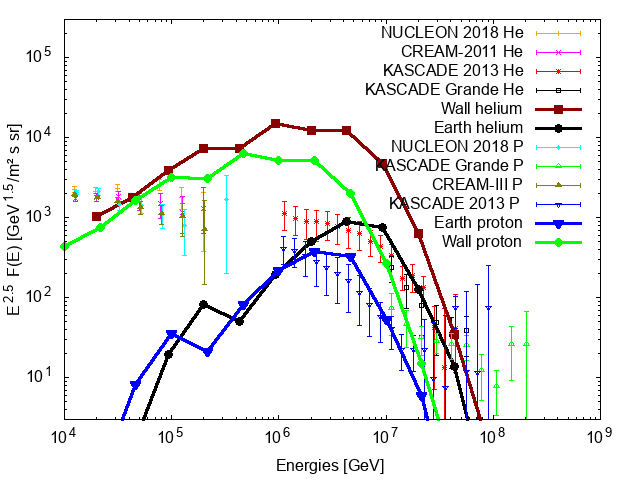}
\includegraphics[width=.45\textwidth]{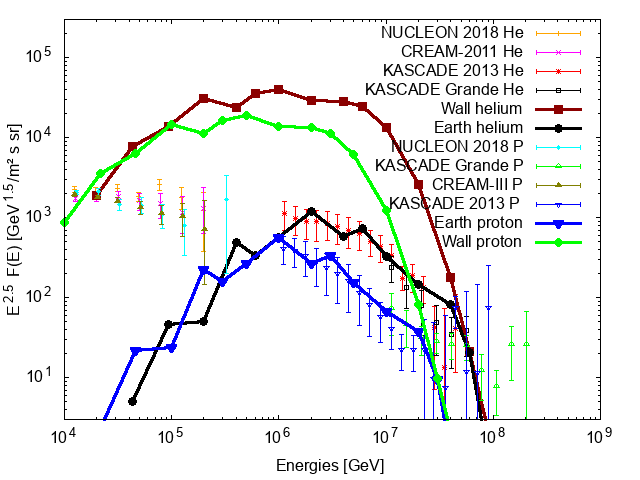}
\end{center}
\caption{Contribution of Vela to the proton (green and blue) and helium
  (dark-red and black lines) fluxes in the bubble
  wall and at Earth compared to the data from
  Refs.~\cite{Gorbunov:2018stf,Yoon:2017qjx,Apel:2013uni}; top panel for
  case U, lower panel for case JF.}
\label{fig:protons_helium}
\end{figure}

In Fig.~\ref{fig:protons_helium}, we compare the proton and helium fluxes
from Vela in the bubble wall and at Earth compared to the data from
Refs.~\cite{Gorbunov:2018stf,Yoon:2017qjx,Apel:2013uni}.
We see again that the larger transition width, $w_1 = 1$\,pc used
in case JF allows low-rigidity particles to enter more easily the
bubble. As a result, a larger fraction of low-energy data can be
explained by the contribution from Vela. Note also that the flux
is dominated by helium in the region most interesting for the
secondary production of neutrinos and photons.

\begin{figure}
  \centering
  \includegraphics[width=0.45\textwidth]{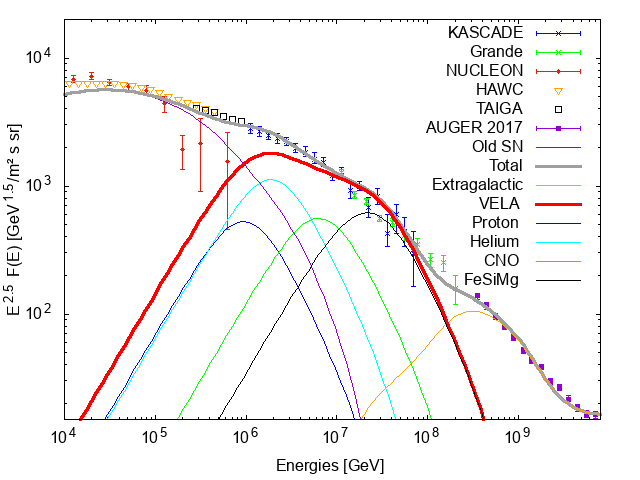}
  \includegraphics[width=0.45\textwidth]{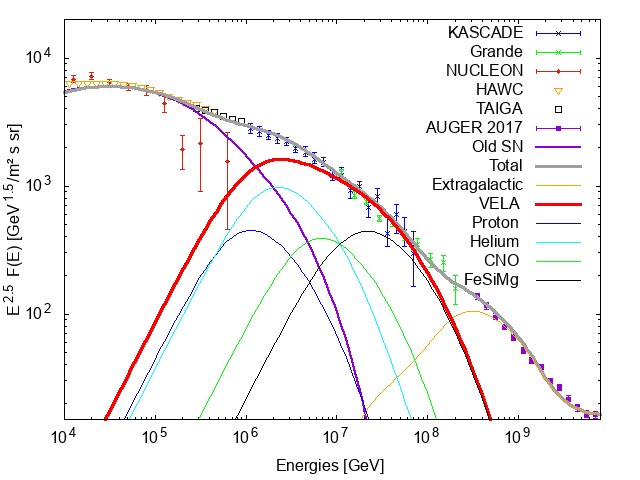}
  \caption{The all-particles flux from Vela, from a 2--3\,Myr
    SN~\cite{Kachelriess:2017yzq}  and the
  extragalactic contribution from Ref.~\cite{Kachelriess:2017tvs} together with
  experimental data from NUCLEON~\cite{Gorbunov:2018}, HAWC~\cite{Alfaro:2017cwx}, TAIGA~\cite{Berezhnev:2015zga}, CREAM~\cite{Yoon:2011cr}, KASCADE and KASCADE~Grande \cite{Apel:2013uni}, and AUGER~\cite{Fenu:2017hlc}; top panel for
  case U, lower panel for case JF.
  \label{fig:all}}
\end{figure}

From Fig.~\ref{fig:all}, we see that the all-particles flux fits well the
experimental data up to $10^{17}$\,eV.
In the energy range above $10^{17}$\,eV, the extragalactic contribution
becomes important which we model following
Ref.~\cite{Kachelriess:2017tvs}. 
We compute the total energy output of Vela and the relative contribution
of the different nuclear groups from the normalisation of the
simulated data to the experimental ones. In the case JF, the relative energy
fraction in protons found is~0.54, the one of helium~0.42, of CNO~ 0.03 and
of \mbox{FeSiMg}~0.007, respectively. We obtain then as total energy output in 
CRs $4.2 \times 10^{49}$\,erg. In the case U, the relative energy fraction
in protons found is~0.55, the one of helium~0.42, of CNO~0.025 and
of \mbox{FeSiMg}~0.004, respectively, and the total energy output in 
CRs $3.6 \times 10^{49}$\,erg. The total kinetic energy of the Vela supernova
calculated in Ref.~\cite{Sushch:2010je} is $1.4 \times 10^{50}$\,erg.
 We note also that
the CR acceleration efficiency of Vela should be high, as it is expected
in the scenario of strong magnetic field amplification of
Refs.~\cite{2001MNRAS.321..433B,2004MNRAS.353..550B}.

\subsection{Neutrino and photon fluxes}
\label{neutrino}

The CR flux in the knee region is dominated by helium. Since the concept
of a ``nuclear enhancement factor''  is not well
defined~\cite{Kachelriess:2014mga,Kachelriess:2019}, 
we employ the Monte Carlo generator
QGSJET-II~\citep{Ostapchenko:2006nh,Ostapchenko:2010vb} 
to calculate the photon and neutrino secondary fluxes. 
We assume a mass fraction of 24\% of Helium in the target
gas and calculate the average intensity of the
secondaries as
\begin{align}
 I_i(E) & = \frac{c}{4\pi} \sum_{A,A'\in\{1,4\}} \int_E^\infty{\rm d}E' \,
          \frac{{\rm d}\sigma_{\rm inel}^{AA'\to i}(E',E)}{{\rm d}E}
\\ & \times
          \int{\rm d}^3x \,  \frac{n_A(E',\vec x) n_{\rm gas}^{A'}(\vec x)}{d^2}\,,
\end{align}
where $\sigma_{\rm inel}^{AA'}$ is the production cross section of secondaries
of type $i$ in interactions of nuclei with mass number $A$ and $A'$,
$d$ denotes the distance from the Sun to the interaction
point $\vec x$, $n_p(E,\vec x)$ the differential number density
of CR protons and $n_{\rm gas}^A(\vec x)$  the density of protons
and Helium in the bubble wall. We approximate the CR density
$n_A(E,\vec x)$ by the average CR density in the wall calculated
previously.

\begin{figure}
  \centering
\includegraphics[width=.45\textwidth]{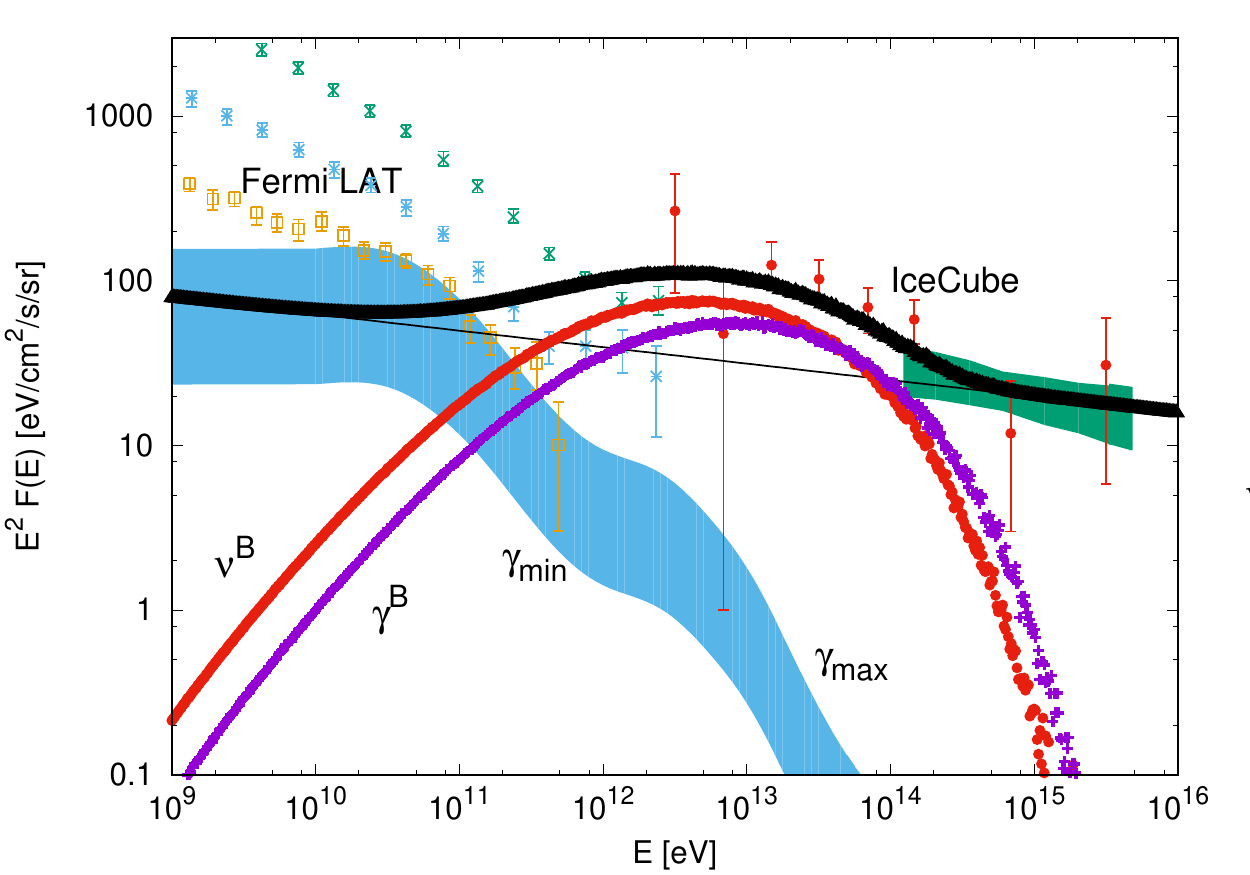}
\includegraphics[width=.45\textwidth]{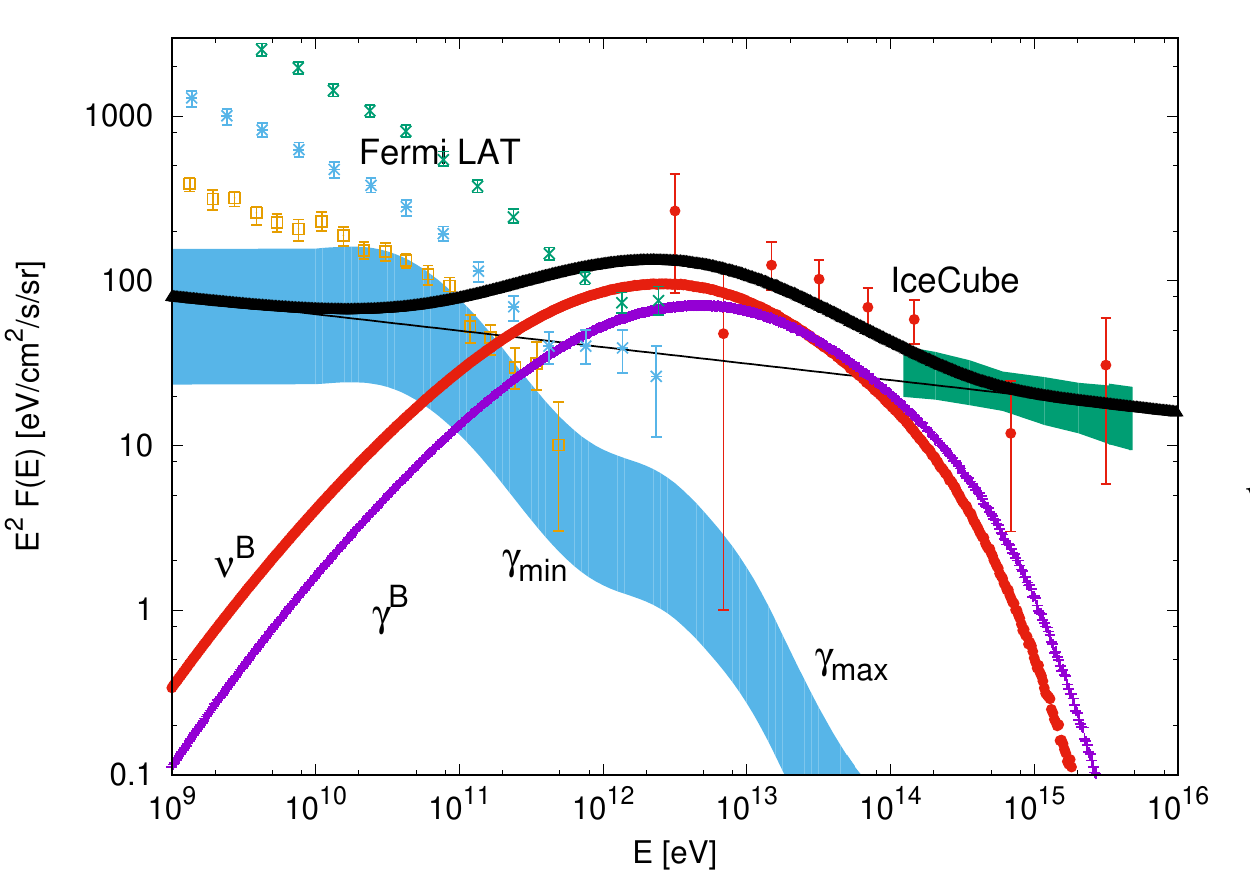}
\caption{Multi-messenger contribution  in neutrinos and gamma-rays of  cosmic ray interactions in the walls of Local Bubble compared to Fermi LAT and IceCube measurements.
Simulations of neutrino and gamma-ray fluxes from Local Bubble in our model are presented with red circles and magenta crosses. 
Neutrino flux from extragalactic sources and total neutrino flux including contribution of bubble are presented with  black lines for $1/E^{2.1}$. 
 Corresponding diffuse gamma-ray flux from extragalactic sources give contribution within blue strip normalized to 
diffuse gamma-ray background measured by Fermi Lat, presented with orange error-bars. Average diffuse gamma-ray flux at high galactic latitudes 
$|b|>20^\circ$ is presented with blue points and middle Galactic latitude flux $10^\circ<|b|<30^\circ$ with green points. At top panel we present case of model with simplified GMF outside of bubble and in bottom panel for GF12 model.}
\label{fig:nu}
\end{figure}

In Fig.~\ref{fig:nu} we compare the flux of neutrinos (red circles) and
gamma-rays (magenta crosses) produced by CR interactions in the wall of the
Local Bubble to Fermi-LAT and IceCube measurements.
The IceCube neutrino data consist of the muon neutrino channel with
measurements above the atmospheric background at $E>100$\,TeV (green band)
and cascade events which show an excess with respect to the continuation of
the muon neutrino flux at $E<100$ TeV (red data with errorbars).
Additionally, we show the neutrino flux from extragalactic sources 
as a thin black line for a  $1/E^{2.1}$ power law, and the sum of the Galactic
and extragalactic contributions with a thick black line. One can see that
the sum of the two neutrino fluxes well fits the IceCube data.

\begin{figure}[t]
  \centering
\includegraphics[width=.45\textwidth]{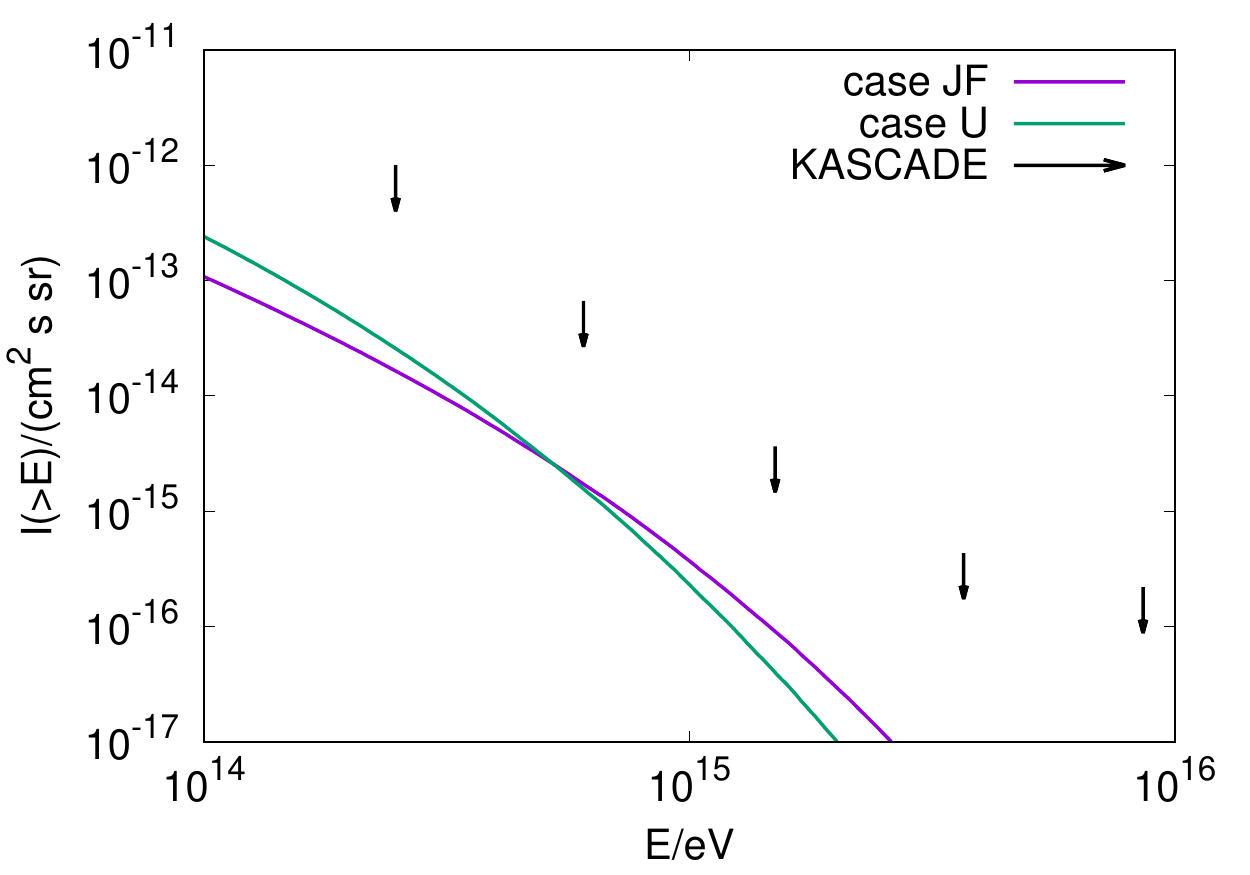}
\caption{Integral photon flux from cosmic ray interactions in the walls of Local Bubble compared to KASCADE limits.}
\label{fig:photons}
\end{figure}

The extragalactic diffuse gamma-ray flux measured by Fermi-Lat is shown
with orange errorbars together with estimates for the contributions to
the gamma-ray flux from extragalactic neutrino sources. The minimal flux
close to the lowest values in the blue band corresponds to an $1/E^2$ neutrino
flux, while the highest values is normalized to the  measured diffuse
gamma-ray background. Since up to $85 \%$ of this background comes from
unresolved blazars, the contribution of all other sources is restricted to
the lowest part of the blue band, corresponding to an $1/E^2$ neutrino flux.
The average diffuse gamma-ray flux at high galactic latitudes 
$|b|>20^\circ$ is presented with green points. This flux is dominated by the
diffuse emission in the local part of our Galaxy. In addition to the
expected cutoff in the diffuse emission above 100\,GeV one can see a 
new hard component at $E>300$ GeV, found in Ref.~\cite{Neronov:2018ibl}.
This component is fitted with the gamma-ray flux from the walls of
the Local Bubble. The neutrino counterpart extends also to $|b|>20^\circ$, 
and has a large contribution from high Galactic latitudes.

The integral photon flux from CR interactions in the wall of the
LB are compared to KASCADE limits in Fig.~\ref{fig:photons}. In order to
detect this photon flux, new more sensitive experiments are required.
The flux level of gamma-rays predicted in our model at PeV energies may
be measured by CR experiments with a high hadron rejection power,
like CARPET-3~\cite{Dzhappuev:2018bnl} or LHAASO~\cite{DiSciascio:2016rgi}.

\section{Conclusions}
\label{concl}

In this work, we presented a model which explains the high diffuse neutrino
flux measured by IceCube at  $E<100$ TeV by CR interactions in the
walls of the Local Bubble. One of the main ingredients of our model is the
presence of a young, nearby CR source like the Vela SNR. In typical models
of the large-scale regular Galactic magnetic field like the one of
Jansson and Farrar, where the Sun and Vela are connected by a regular magnetic
field line, the resulting CR flux on Earth would overshot the observed level.

The strong magnetic field in the wall of the LB serves as a magnetic
shield for CRs from such sources. In particular, the CR flux in the wall is
much higher compared to the one measured on Earth.
At the same time, the gas density is $\sim 10$ times higher in the wall
compared to the interstellar one. Both factors lead to a significant neutrino
and gamma-ray flux produced in the bubble walls. Combining this Galactic
component with a standard $1/E^{2.1}$ extragalactic neutrino flux describes
well the IceCube neutrino spectrum. The accompanying photon flux may be
responsible for the TeV gamma-ray excess found recently in
Ref.~\cite{Neronov:2018ibl}.  This gamma-ray flux can be detected by
gamma-ray sensitive CR experiments like LHAASO or Carpet-3.
For instance, the Carpet-3 experiment will be sensitive in the sub-PeV
energy range. At lower energies, a next generation gamma-ray experiment
which is more sensitive than Fermi-LAT would be required to detect these
gamma-rays.

\acknowledgments
\noindent
MK would like to thank the theory group at APC for hospitality.


\begin{thebibliography}{59}
\expandafter\ifx\csname natexlab\endcsname\relax\def\natexlab#1{#1}\fi
\expandafter\ifx\csname bibnamefont\endcsname\relax
  \def\bibnamefont#1{#1}\fi
\expandafter\ifx\csname bibfnamefont\endcsname\relax
  \def\bibfnamefont#1{#1}\fi
\expandafter\ifx\csname citenamefont\endcsname\relax
  \def\citenamefont#1{#1}\fi
\expandafter\ifx\csname url\endcsname\relax
  \def\url#1{\texttt{#1}}\fi
\expandafter\ifx\csname urlprefix\endcsname\relax\def\urlprefix{URL }\fi
\providecommand{\bibinfo}[2]{#2}
\providecommand{\eprint}[2][]{\url{#2}}

\bibitem[{\citenamefont{Gaisser et~al.}(1995)\citenamefont{Gaisser, Halzen, and
  Stanev}}]{Gaisser:1994yf}
\bibinfo{author}{\bibfnamefont{T.~K.} \bibnamefont{Gaisser}},
  \bibinfo{author}{\bibfnamefont{F.}~\bibnamefont{Halzen}}, \bibnamefont{and}
  \bibinfo{author}{\bibfnamefont{T.}~\bibnamefont{Stanev}},
  \bibinfo{journal}{Phys. Rept.} \textbf{\bibinfo{volume}{258}},
  \bibinfo{pages}{173} (\bibinfo{year}{1995}), \bibinfo{note}{[Erratum: Phys.
  Rept.271,355(1996)]}, \eprint{hep-ph/9410384}.

\bibitem[{\citenamefont{Aartsen et~al.}(2016)}]{Aartsen:2016xlq}
\bibinfo{author}{\bibfnamefont{M.~G.} \bibnamefont{Aartsen}}
  \bibnamefont{et~al.} (\bibinfo{collaboration}{IceCube}),
  \bibinfo{journal}{Astrophys. J.} \textbf{\bibinfo{volume}{833}},
  \bibinfo{pages}{3} (\bibinfo{year}{2016}), \eprint{1607.08006}.

\bibitem[{\citenamefont{Aartsen et~al.}(2020)}]{Aartsen:2020aqd}
\bibinfo{author}{\bibfnamefont{M.~G.} \bibnamefont{Aartsen}}
  \bibnamefont{et~al.} (\bibinfo{collaboration}{IceCube})
  (\bibinfo{year}{2020}), \eprint{2001.09520}.

\bibitem[{\citenamefont{Stecker et~al.}(1991)\citenamefont{Stecker, Done,
  Salamon, and Sommers}}]{Stecker:1991vm}
\bibinfo{author}{\bibfnamefont{F.~W.} \bibnamefont{Stecker}},
  \bibinfo{author}{\bibfnamefont{C.}~\bibnamefont{Done}},
  \bibinfo{author}{\bibfnamefont{M.~H.} \bibnamefont{Salamon}},
  \bibnamefont{and} \bibinfo{author}{\bibfnamefont{P.}~\bibnamefont{Sommers}},
  \bibinfo{journal}{Phys. Rev. Lett.} \textbf{\bibinfo{volume}{66}},
  \bibinfo{pages}{2697} (\bibinfo{year}{1991}), \bibinfo{note}{[Erratum: Phys.
  Rev. Lett.69,2738(1992)]}.

\bibitem[{\citenamefont{Mannheim}(1995)}]{Mannheim:1995mm}
\bibinfo{author}{\bibfnamefont{K.}~\bibnamefont{Mannheim}},
  \bibinfo{journal}{Astropart. Phys.} \textbf{\bibinfo{volume}{3}},
  \bibinfo{pages}{295} (\bibinfo{year}{1995}).

\bibitem[{\citenamefont{Waxman and Bahcall}(1997)}]{Waxman:1997ti}
\bibinfo{author}{\bibfnamefont{E.}~\bibnamefont{Waxman}} \bibnamefont{and}
  \bibinfo{author}{\bibfnamefont{J.~N.} \bibnamefont{Bahcall}},
  \bibinfo{journal}{Phys. Rev. Lett.} \textbf{\bibinfo{volume}{78}},
  \bibinfo{pages}{2292} (\bibinfo{year}{1997}), \eprint{astro-ph/9701231}.

\bibitem[{\citenamefont{Loeb and Waxman}(2006)}]{Loeb:2006tw}
\bibinfo{author}{\bibfnamefont{A.}~\bibnamefont{Loeb}} \bibnamefont{and}
  \bibinfo{author}{\bibfnamefont{E.}~\bibnamefont{Waxman}},
  \bibinfo{journal}{JCAP} \textbf{\bibinfo{volume}{0605}}, \bibinfo{pages}{003}
  (\bibinfo{year}{2006}), \eprint{astro-ph/0601695}.

\bibitem[{\citenamefont{Berezinsky et~al.}(2011)\citenamefont{Berezinsky,
  Gazizov, Kachelrie{\ss}, and Ostapchenko}}]{Berezinsky:2010xa}
\bibinfo{author}{\bibfnamefont{V.}~\bibnamefont{Berezinsky}},
  \bibinfo{author}{\bibfnamefont{A.}~\bibnamefont{Gazizov}},
  \bibinfo{author}{\bibfnamefont{M.}~\bibnamefont{Kachelrie{\ss}}},
  \bibnamefont{and}
  \bibinfo{author}{\bibfnamefont{S.}~\bibnamefont{Ostapchenko}},
  \bibinfo{journal}{Phys. Lett.} \textbf{\bibinfo{volume}{B695}},
  \bibinfo{pages}{13} (\bibinfo{year}{2011}), \eprint{1003.1496}.

\bibitem[{\citenamefont{Murase et~al.}(2013)\citenamefont{Murase, Ahlers, and
  Lacki}}]{Murase:2013rfa}
\bibinfo{author}{\bibfnamefont{K.}~\bibnamefont{Murase}},
  \bibinfo{author}{\bibfnamefont{M.}~\bibnamefont{Ahlers}}, \bibnamefont{and}
  \bibinfo{author}{\bibfnamefont{B.~C.} \bibnamefont{Lacki}},
  \bibinfo{journal}{Phys. Rev.} \textbf{\bibinfo{volume}{D88}},
  \bibinfo{pages}{121301} (\bibinfo{year}{2013}), \eprint{1306.3417}.

\bibitem[{\citenamefont{Neronov and
  Semikoz}(2016{\natexlab{a}})}]{Neronov:2014uma}
\bibinfo{author}{\bibfnamefont{A.}~\bibnamefont{Neronov}} \bibnamefont{and}
  \bibinfo{author}{\bibfnamefont{D.~V.} \bibnamefont{Semikoz}},
  \bibinfo{journal}{Astropart. Phys.} \textbf{\bibinfo{volume}{72}},
  \bibinfo{pages}{32} (\bibinfo{year}{2016}{\natexlab{a}}), \eprint{1412.1690}.

\bibitem[{\citenamefont{Albert et~al.}(2017)}]{Albert:2017oba}
\bibinfo{author}{\bibfnamefont{A.}~\bibnamefont{Albert}} \bibnamefont{et~al.}
  (\bibinfo{collaboration}{ANTARES}), \bibinfo{journal}{Phys. Rev.}
  \textbf{\bibinfo{volume}{D96}}, \bibinfo{pages}{062001}
  (\bibinfo{year}{2017}), \eprint{1705.00497}.

\bibitem[{\citenamefont{Neronov and
  Semikoz}(2016{\natexlab{b}})}]{Neronov:2015osa}
\bibinfo{author}{\bibfnamefont{A.}~\bibnamefont{Neronov}} \bibnamefont{and}
  \bibinfo{author}{\bibfnamefont{D.~V.} \bibnamefont{Semikoz}},
  \bibinfo{journal}{Astropart. Phys.} \textbf{\bibinfo{volume}{75}},
  \bibinfo{pages}{60} (\bibinfo{year}{2016}{\natexlab{b}}),
  \eprint{1509.03522}.

\bibitem[{\citenamefont{Neronov and
  Semikoz}(2016{\natexlab{c}})}]{Neronov:2016bnp}
\bibinfo{author}{\bibfnamefont{A.}~\bibnamefont{Neronov}} \bibnamefont{and}
  \bibinfo{author}{\bibfnamefont{D.~V.} \bibnamefont{Semikoz}},
  \bibinfo{journal}{Phys. Rev.} \textbf{\bibinfo{volume}{D93}},
  \bibinfo{pages}{123002} (\bibinfo{year}{2016}{\natexlab{c}}),
  \eprint{1603.06733}.

\bibitem[{\citenamefont{Palladino et~al.}(2017)\citenamefont{Palladino,
  Mascaretti, and Vissani}}]{Palladino:2017qda}
\bibinfo{author}{\bibfnamefont{A.}~\bibnamefont{Palladino}},
  \bibinfo{author}{\bibfnamefont{C.}~\bibnamefont{Mascaretti}},
  \bibnamefont{and} \bibinfo{author}{\bibfnamefont{F.}~\bibnamefont{Vissani}},
  \bibinfo{journal}{Eur. Phys. J.} \textbf{\bibinfo{volume}{C77}},
  \bibinfo{pages}{684} (\bibinfo{year}{2017}), \eprint{1708.02094}.

\bibitem[{\citenamefont{Palladino and Winter}(2018)}]{Palladino:2018evm}
\bibinfo{author}{\bibfnamefont{A.}~\bibnamefont{Palladino}} \bibnamefont{and}
  \bibinfo{author}{\bibfnamefont{W.}~\bibnamefont{Winter}}
  (\bibinfo{year}{2018}), \eprint{1801.07277}.

\bibitem[{\citenamefont{Neronov et~al.}(2018)\citenamefont{Neronov,
  Kachelrie{\ss}, and Semikoz}}]{Neronov:2018ibl}
\bibinfo{author}{\bibfnamefont{A.}~\bibnamefont{Neronov}},
  \bibinfo{author}{\bibfnamefont{M.}~\bibnamefont{Kachelrie{\ss}}},
  \bibnamefont{and} \bibinfo{author}{\bibfnamefont{D.~V.}
  \bibnamefont{Semikoz}}, \bibinfo{journal}{Phys. Rev.}
  \textbf{\bibinfo{volume}{D98}}, \bibinfo{pages}{023004}
  (\bibinfo{year}{2018}), \eprint{1802.09983}.

\bibitem[{\citenamefont{Neronov and Semikoz}(2019)}]{Neronov:2019ncc}
\bibinfo{author}{\bibfnamefont{A.}~\bibnamefont{Neronov}} \bibnamefont{and}
  \bibinfo{author}{\bibfnamefont{D.}~\bibnamefont{Semikoz}}
  (\bibinfo{year}{2019}), \eprint{1907.06061}.

\bibitem[{\citenamefont{Taylor et~al.}(2014)\citenamefont{Taylor, Gabici, and
  Aharonian}}]{Taylor:2014hya}
\bibinfo{author}{\bibfnamefont{A.~M.} \bibnamefont{Taylor}},
  \bibinfo{author}{\bibfnamefont{S.}~\bibnamefont{Gabici}}, \bibnamefont{and}
  \bibinfo{author}{\bibfnamefont{F.}~\bibnamefont{Aharonian}},
  \bibinfo{journal}{Phys. Rev.} \textbf{\bibinfo{volume}{D89}},
  \bibinfo{pages}{103003} (\bibinfo{year}{2014}), \eprint{1403.3206}.

\bibitem[{\citenamefont{Blasi and Amato}(2019)}]{Blasi:2019obb}
\bibinfo{author}{\bibfnamefont{P.}~\bibnamefont{Blasi}} \bibnamefont{and}
  \bibinfo{author}{\bibfnamefont{E.}~\bibnamefont{Amato}},
  \bibinfo{journal}{Phys. Rev. Lett.} \textbf{\bibinfo{volume}{122}},
  \bibinfo{pages}{051101} (\bibinfo{year}{2019}), \eprint{1901.03609}.

\bibitem[{\citenamefont{Kalashev and Troitsky}(2016)}]{Kalashev:2016euk}
\bibinfo{author}{\bibfnamefont{O.}~\bibnamefont{Kalashev}} \bibnamefont{and}
  \bibinfo{author}{\bibfnamefont{S.}~\bibnamefont{Troitsky}},
  \bibinfo{journal}{Phys. Rev. D} \textbf{\bibinfo{volume}{94}},
  \bibinfo{pages}{063013} (\bibinfo{year}{2016}), \eprint{1608.07421}.

\bibitem[{\citenamefont{Berezinsky et~al.}(1997)\citenamefont{Berezinsky,
  Kachelrie{\ss}, and Vilenkin}}]{Berezinsky:1997hy}
\bibinfo{author}{\bibfnamefont{V.}~\bibnamefont{Berezinsky}},
  \bibinfo{author}{\bibfnamefont{M.}~\bibnamefont{Kachelrie{\ss}}},
  \bibnamefont{and} \bibinfo{author}{\bibfnamefont{A.}~\bibnamefont{Vilenkin}},
  \bibinfo{journal}{Phys. Rev. Lett.} \textbf{\bibinfo{volume}{79}},
  \bibinfo{pages}{4302} (\bibinfo{year}{1997}), \eprint{astro-ph/9708217}.

\bibitem[{\citenamefont{Feldstein et~al.}(2013)\citenamefont{Feldstein,
  Kusenko, Matsumoto, and Yanagida}}]{Feldstein:2013kka}
\bibinfo{author}{\bibfnamefont{B.}~\bibnamefont{Feldstein}},
  \bibinfo{author}{\bibfnamefont{A.}~\bibnamefont{Kusenko}},
  \bibinfo{author}{\bibfnamefont{S.}~\bibnamefont{Matsumoto}},
  \bibnamefont{and} \bibinfo{author}{\bibfnamefont{T.~T.}
  \bibnamefont{Yanagida}}, \bibinfo{journal}{Phys. Rev.}
  \textbf{\bibinfo{volume}{D88}}, \bibinfo{pages}{015004}
  (\bibinfo{year}{2013}), \eprint{1303.7320}.

\bibitem[{\citenamefont{Esmaili and Serpico}(2013)}]{Esmaili:2013gha}
\bibinfo{author}{\bibfnamefont{A.}~\bibnamefont{Esmaili}} \bibnamefont{and}
  \bibinfo{author}{\bibfnamefont{P.~D.} \bibnamefont{Serpico}},
  \bibinfo{journal}{JCAP} \textbf{\bibinfo{volume}{1311}}, \bibinfo{pages}{054}
  (\bibinfo{year}{2013}), \eprint{1308.1105}.

\bibitem[{\citenamefont{Crocker and Aharonian}(2011)}]{Crocker:2010dg}
\bibinfo{author}{\bibfnamefont{R.~M.} \bibnamefont{Crocker}} \bibnamefont{and}
  \bibinfo{author}{\bibfnamefont{F.}~\bibnamefont{Aharonian}},
  \bibinfo{journal}{Phys. Rev. Lett.} \textbf{\bibinfo{volume}{106}},
  \bibinfo{pages}{101102} (\bibinfo{year}{2011}), \eprint{1008.2658}.

\bibitem[{\citenamefont{Lunardini and Razzaque}(2012)}]{Lunardini:2011br}
\bibinfo{author}{\bibfnamefont{C.}~\bibnamefont{Lunardini}} \bibnamefont{and}
  \bibinfo{author}{\bibfnamefont{S.}~\bibnamefont{Razzaque}},
  \bibinfo{journal}{Phys. Rev. Lett.} \textbf{\bibinfo{volume}{108}},
  \bibinfo{pages}{221102} (\bibinfo{year}{2012}), \eprint{1112.4799}.

\bibitem[{\citenamefont{Andersen et~al.}(2018)\citenamefont{Andersen,
  Kachelrie{\ss}, and Semikoz}}]{Andersen:2017yyg}
\bibinfo{author}{\bibfnamefont{K.~J.} \bibnamefont{Andersen}},
  \bibinfo{author}{\bibfnamefont{M.}~\bibnamefont{Kachelrie{\ss}}},
  \bibnamefont{and} \bibinfo{author}{\bibfnamefont{D.~V.}
  \bibnamefont{Semikoz}}, \bibinfo{journal}{Astrophys. J.}
  \textbf{\bibinfo{volume}{861}}, \bibinfo{pages}{L19} (\bibinfo{year}{2018}),
  \eprint{1712.03153}.

\bibitem[{\citenamefont{Dzhappuev et~al.}(2019)}]{Dzhappuev:2018bnl}
\bibinfo{author}{\bibfnamefont{D.~D.} \bibnamefont{Dzhappuev}}
  \bibnamefont{et~al.}, \bibinfo{journal}{EPJ Web Conf.}
  \textbf{\bibinfo{volume}{207}}, \bibinfo{pages}{03004}
  (\bibinfo{year}{2019}), \eprint{1812.02663}.

\bibitem[{\citenamefont{Di~Sciascio}(2016)}]{DiSciascio:2016rgi}
\bibinfo{author}{\bibfnamefont{G.}~\bibnamefont{Di~Sciascio}}
  (\bibinfo{collaboration}{LHAASO}), \bibinfo{journal}{Nucl. Part. Phys. Proc.}
  \textbf{\bibinfo{volume}{279-281}}, \bibinfo{pages}{166}
  (\bibinfo{year}{2016}), \eprint{1602.07600}.

\bibitem[{\citenamefont{Desiati}(2014)}]{Desiati:2013lea}
\bibinfo{author}{\bibfnamefont{P.}~\bibnamefont{Desiati}}
  (\bibinfo{collaboration}{IceCube}), \bibinfo{journal}{Nucl. Instrum. Meth. A}
  \textbf{\bibinfo{volume}{742}}, \bibinfo{pages}{199} (\bibinfo{year}{2014}),
  \eprint{1308.0246}.

\bibitem[{\citenamefont{Bouyahiaoui et~al.}(2019)\citenamefont{Bouyahiaoui,
  Kachelrie\ss, and Semikoz}}]{Bouyahiaoui:2018lew}
\bibinfo{author}{\bibfnamefont{M.}~\bibnamefont{Bouyahiaoui}},
  \bibinfo{author}{\bibfnamefont{M.}~\bibnamefont{Kachelrie\ss}},
  \bibnamefont{and} \bibinfo{author}{\bibfnamefont{D.~V.}
  \bibnamefont{Semikoz}}, \bibinfo{journal}{JCAP}
  \textbf{\bibinfo{volume}{1901}}, \bibinfo{pages}{046} (\bibinfo{year}{2019}),
  \eprint{1812.03522}.

\bibitem[{\citenamefont{Jansson and Farrar}(2012)}]{Jansson:2012rt}
\bibinfo{author}{\bibfnamefont{R.}~\bibnamefont{Jansson}} \bibnamefont{and}
  \bibinfo{author}{\bibfnamefont{G.~R.} \bibnamefont{Farrar}},
  \bibinfo{journal}{Astrophys.J.} \textbf{\bibinfo{volume}{761}},
  \bibinfo{pages}{L11} (\bibinfo{year}{2012}), \eprint{1210.7820}.

\bibitem[{\citenamefont{Giacinti et~al.}(2014)\citenamefont{Giacinti,
  Kachelrie{\ss}, and Semikoz}}]{Giacinti:2014xya}
\bibinfo{author}{\bibfnamefont{G.}~\bibnamefont{Giacinti}},
  \bibinfo{author}{\bibfnamefont{M.}~\bibnamefont{Kachelrie{\ss}}},
  \bibnamefont{and} \bibinfo{author}{\bibfnamefont{D.~V.}
  \bibnamefont{Semikoz}}, \bibinfo{journal}{Phys. Rev.}
  \textbf{\bibinfo{volume}{D90}}, \bibinfo{pages}{041302}
  (\bibinfo{year}{2014}), \eprint{1403.3380}.

\bibitem[{\citenamefont{Giacinti et~al.}(2015)\citenamefont{Giacinti,
  Kachelrie{\ss}, and Semikoz}}]{Giacinti:2015hva}
\bibinfo{author}{\bibfnamefont{G.}~\bibnamefont{Giacinti}},
  \bibinfo{author}{\bibfnamefont{M.}~\bibnamefont{Kachelrie{\ss}}},
  \bibnamefont{and} \bibinfo{author}{\bibfnamefont{D.~V.}
  \bibnamefont{Semikoz}}, \bibinfo{journal}{Phys. Rev.}
  \textbf{\bibinfo{volume}{D91}}, \bibinfo{pages}{083009}
  (\bibinfo{year}{2015}), \eprint{1502.01608}.

\bibitem[{\citenamefont{Welsh and Shelton}(2009)}]{Welsh:2009sg}
\bibinfo{author}{\bibfnamefont{B.~Y.} \bibnamefont{Welsh}} \bibnamefont{and}
  \bibinfo{author}{\bibfnamefont{R.~L.} \bibnamefont{Shelton}},
  \bibinfo{journal}{Astrophys. Space Sci.} \textbf{\bibinfo{volume}{323}},
  \bibinfo{pages}{1} (\bibinfo{year}{2009}), \eprint{0906.2827}.

\bibitem[{\citenamefont{Giacinti et~al.}(2012)\citenamefont{Giacinti,
  Kachelrie{\ss}, Semikoz, and Sigl}}]{Giacinti:2011ww}
\bibinfo{author}{\bibfnamefont{G.}~\bibnamefont{Giacinti}},
  \bibinfo{author}{\bibfnamefont{M.}~\bibnamefont{Kachelrie{\ss}}},
  \bibinfo{author}{\bibfnamefont{D.~V.} \bibnamefont{Semikoz}},
  \bibnamefont{and} \bibinfo{author}{\bibfnamefont{G.}~\bibnamefont{Sigl}},
  \bibinfo{journal}{JCAP} \textbf{\bibinfo{volume}{1207}}, \bibinfo{pages}{031}
  (\bibinfo{year}{2012}), \eprint{1112.5599}.

\bibitem[{\citenamefont{Giacinti et~al.}(2018)\citenamefont{Giacinti,
  Kachelrie{\ss}, and Semikoz}}]{Giacinti:2017dgt}
\bibinfo{author}{\bibfnamefont{G.}~\bibnamefont{Giacinti}},
  \bibinfo{author}{\bibfnamefont{M.}~\bibnamefont{Kachelrie{\ss}}},
  \bibnamefont{and} \bibinfo{author}{\bibfnamefont{D.~V.}
  \bibnamefont{Semikoz}}, \bibinfo{journal}{JCAP}
  \textbf{\bibinfo{volume}{1807}}, \bibinfo{pages}{051} (\bibinfo{year}{2018}),
  \eprint{1710.08205}.

\bibitem[{\citenamefont{Knie et~al.}(1999)\citenamefont{Knie, Korschinek,
  Faestermann, Wallner, Scholten et~al.}}]{Knie:1999zz}
\bibinfo{author}{\bibfnamefont{K.}~\bibnamefont{Knie}},
  \bibinfo{author}{\bibfnamefont{G.}~\bibnamefont{Korschinek}},
  \bibinfo{author}{\bibfnamefont{T.}~\bibnamefont{Faestermann}},
  \bibinfo{author}{\bibfnamefont{C.}~\bibnamefont{Wallner}},
  \bibinfo{author}{\bibfnamefont{J.}~\bibnamefont{Scholten}},
  \bibnamefont{et~al.}, \bibinfo{journal}{Phys.Rev.Lett.}
  \textbf{\bibinfo{volume}{83}}, \bibinfo{pages}{18} (\bibinfo{year}{1999}).

\bibitem[{\citenamefont{Benitez et~al.}(2002)\citenamefont{Benitez,
  Maiz-Apellaniz, and Canelles}}]{Benitez:2002jt}
\bibinfo{author}{\bibfnamefont{N.}~\bibnamefont{Benitez}},
  \bibinfo{author}{\bibfnamefont{J.}~\bibnamefont{Maiz-Apellaniz}},
  \bibnamefont{and} \bibinfo{author}{\bibfnamefont{M.}~\bibnamefont{Canelles}},
  \bibinfo{journal}{Phys. Rev. Lett.} \textbf{\bibinfo{volume}{88}},
  \bibinfo{pages}{081101} (\bibinfo{year}{2002}), \eprint{astro-ph/0201018}.

\bibitem[{\citenamefont{Fitoussi et~al.}(2008)}]{Fitoussi:2007ef}
\bibinfo{author}{\bibfnamefont{C.}~\bibnamefont{Fitoussi}}
  \bibnamefont{et~al.}, \bibinfo{journal}{Phys. Rev. Lett.}
  \textbf{\bibinfo{volume}{101}}, \bibinfo{pages}{121101}
  (\bibinfo{year}{2008}), \eprint{0709.4197}.

\bibitem[{\citenamefont{{Wallner} et~al.}(2016)\citenamefont{{Wallner},
  {Feige}, {Kinoshita}, {Paul}, {Fifield}, {Golser}, {Honda}, {Linnemann},
  {Matsuzaki}, {Merchel} et~al.}}]{2016Natur.532...69W}
\bibinfo{author}{\bibfnamefont{A.}~\bibnamefont{{Wallner}}},
  \bibinfo{author}{\bibfnamefont{J.}~\bibnamefont{{Feige}}},
  \bibinfo{author}{\bibfnamefont{N.}~\bibnamefont{{Kinoshita}}},
  \bibinfo{author}{\bibfnamefont{M.}~\bibnamefont{{Paul}}},
  \bibinfo{author}{\bibfnamefont{L.~K.} \bibnamefont{{Fifield}}},
  \bibinfo{author}{\bibfnamefont{R.}~\bibnamefont{{Golser}}},
  \bibinfo{author}{\bibfnamefont{M.}~\bibnamefont{{Honda}}},
  \bibinfo{author}{\bibfnamefont{U.}~\bibnamefont{{Linnemann}}},
  \bibinfo{author}{\bibfnamefont{H.}~\bibnamefont{{Matsuzaki}}},
  \bibinfo{author}{\bibfnamefont{S.}~\bibnamefont{{Merchel}}},
  \bibnamefont{et~al.}, \bibinfo{journal}{\nat} \textbf{\bibinfo{volume}{532}},
  \bibinfo{pages}{69} (\bibinfo{year}{2016}).

\bibitem[{\citenamefont{Drury et~al.}(2003)\citenamefont{Drury, van~der Swaluw,
  and Carroll}}]{Drury:2003fd}
\bibinfo{author}{\bibfnamefont{L.~O.} \bibnamefont{Drury}},
  \bibinfo{author}{\bibfnamefont{E.}~\bibnamefont{van~der Swaluw}},
  \bibnamefont{and} \bibinfo{author}{\bibfnamefont{O.}~\bibnamefont{Carroll}}
  (\bibinfo{year}{2003}), \eprint{astro-ph/0309820}.

\bibitem[{\citenamefont{{Bell} and {Lucek}}(2001)}]{2001MNRAS.321..433B}
\bibinfo{author}{\bibfnamefont{A.~R.} \bibnamefont{{Bell}}} \bibnamefont{and}
  \bibinfo{author}{\bibfnamefont{S.~G.} \bibnamefont{{Lucek}}},
  \bibinfo{journal}{\mnras} \textbf{\bibinfo{volume}{321}},
  \bibinfo{pages}{433} (\bibinfo{year}{2001}).

\bibitem[{\citenamefont{{Bell}}(2004)}]{2004MNRAS.353..550B}
\bibinfo{author}{\bibfnamefont{A.~R.} \bibnamefont{{Bell}}},
  \bibinfo{journal}{\mnras} \textbf{\bibinfo{volume}{353}},
  \bibinfo{pages}{550} (\bibinfo{year}{2004}).

\bibitem[{\citenamefont{Kachelrie{\ss}
  et~al.}(2015)\citenamefont{Kachelrie{\ss}, Neronov, and
  Semikoz}}]{Kachelriess:2015oua}
\bibinfo{author}{\bibfnamefont{M.}~\bibnamefont{Kachelrie{\ss}}},
  \bibinfo{author}{\bibfnamefont{A.}~\bibnamefont{Neronov}}, \bibnamefont{and}
  \bibinfo{author}{\bibfnamefont{D.~V.} \bibnamefont{Semikoz}},
  \bibinfo{journal}{Phys. Rev. Lett.} \textbf{\bibinfo{volume}{115}},
  \bibinfo{pages}{181103} (\bibinfo{year}{2015}), \eprint{1504.06472}.

\bibitem[{\citenamefont{Gorbunov
  et~al.}(2018{\natexlab{a}})}]{Gorbunov:2018stf}
\bibinfo{author}{\bibfnamefont{N.}~\bibnamefont{Gorbunov}} \bibnamefont{et~al.}
  (\bibinfo{year}{2018}{\natexlab{a}}), \eprint{1809.05333}.

\bibitem[{\citenamefont{Yoon et~al.}(2017)}]{Yoon:2017qjx}
\bibinfo{author}{\bibfnamefont{Y.~S.} \bibnamefont{Yoon}} \bibnamefont{et~al.},
  \bibinfo{journal}{Astrophys. J.} \textbf{\bibinfo{volume}{839}},
  \bibinfo{pages}{5} (\bibinfo{year}{2017}), \eprint{1704.02512}.

\bibitem[{\citenamefont{Apel et~al.}(2013)}]{Apel:2013uni}
\bibinfo{author}{\bibfnamefont{W.~D.} \bibnamefont{Apel}} \bibnamefont{et~al.},
  \bibinfo{journal}{Astropart. Phys.} \textbf{\bibinfo{volume}{47}},
  \bibinfo{pages}{54} (\bibinfo{year}{2013}), \eprint{1306.6283}.

\bibitem[{\citenamefont{Kachelrie{\ss}
  et~al.}(2018)\citenamefont{Kachelrie{\ss}, Neronov, and
  Semikoz}}]{Kachelriess:2017yzq}
\bibinfo{author}{\bibfnamefont{M.}~\bibnamefont{Kachelrie{\ss}}},
  \bibinfo{author}{\bibfnamefont{A.}~\bibnamefont{Neronov}}, \bibnamefont{and}
  \bibinfo{author}{\bibfnamefont{D.~V.} \bibnamefont{Semikoz}},
  \bibinfo{journal}{Phys. Rev.} \textbf{\bibinfo{volume}{D97}},
  \bibinfo{pages}{063011} (\bibinfo{year}{2018}), \eprint{1710.02321}.

\bibitem[{\citenamefont{Kachelrie\ss et~al.}(2017)\citenamefont{Kachelrie\ss,
  Kalashev, Ostapchenko, and Semikoz}}]{Kachelriess:2017tvs}
\bibinfo{author}{\bibfnamefont{M.}~\bibnamefont{Kachelrie\ss}},
  \bibinfo{author}{\bibfnamefont{O.}~\bibnamefont{Kalashev}},
  \bibinfo{author}{\bibfnamefont{S.}~\bibnamefont{Ostapchenko}},
  \bibnamefont{and} \bibinfo{author}{\bibfnamefont{D.~V.}
  \bibnamefont{Semikoz}}, \bibinfo{journal}{Phys. Rev.}
  \textbf{\bibinfo{volume}{D96}}, \bibinfo{pages}{083006}
  (\bibinfo{year}{2017}), \eprint{1704.06893}.

\bibitem[{\citenamefont{Gorbunov et~al.}(2018{\natexlab{b}})}]{Gorbunov:2018}
\bibinfo{author}{\bibfnamefont{N.}~\bibnamefont{Gorbunov}} \bibnamefont{et~al.}
  (\bibinfo{year}{2018}{\natexlab{b}}), \eprint{1809.05333}.

\bibitem[{\citenamefont{Alfaro et~al.}(2017)}]{Alfaro:2017cwx}
\bibinfo{author}{\bibfnamefont{R.}~\bibnamefont{Alfaro}} \bibnamefont{et~al.}
  (\bibinfo{collaboration}{HAWC}), \bibinfo{journal}{Phys. Rev.}
  \textbf{\bibinfo{volume}{D96}}, \bibinfo{pages}{122001}
  (\bibinfo{year}{2017}), \eprint{1710.00890}.

\bibitem[{\citenamefont{Berezhnev et~al.}(2015)}]{Berezhnev:2015zga}
\bibinfo{author}{\bibfnamefont{S.~F.} \bibnamefont{Berezhnev}}
  \bibnamefont{et~al.}, \bibinfo{journal}{Bull. Russ. Acad. Sci. Phys.}
  \textbf{\bibinfo{volume}{79}}, \bibinfo{pages}{348} (\bibinfo{year}{2015}),
  \bibinfo{note}{[Izv. Ross. Akad. Nauk Ser. Fiz.79,no.3,381(2015)]}.

\bibitem[{\citenamefont{Yoon et~al.}(2011)}]{Yoon:2011cr}
\bibinfo{author}{\bibfnamefont{Y.~S.} \bibnamefont{Yoon}} \bibnamefont{et~al.},
  \bibinfo{journal}{Astrophys. J.} \textbf{\bibinfo{volume}{728}},
  \bibinfo{pages}{8} (\bibinfo{year}{2011}), \eprint{1602.04710}.

\bibitem[{\citenamefont{Fenu}(2017)}]{Fenu:2017hlc}
\bibinfo{author}{\bibfnamefont{F.}~\bibnamefont{Fenu}}
  (\bibinfo{collaboration}{Pierre Auger}), pp. \bibinfo{pages}{9--16}
  (\bibinfo{year}{2017}), \bibinfo{note}{[PoSICRC2017,486(2018)]}.

\bibitem[{\citenamefont{Sushch et~al.}(2011)\citenamefont{Sushch, Hnatyk, and
  Neronov}}]{Sushch:2010je}
\bibinfo{author}{\bibfnamefont{I.}~\bibnamefont{Sushch}},
  \bibinfo{author}{\bibfnamefont{B.}~\bibnamefont{Hnatyk}}, \bibnamefont{and}
  \bibinfo{author}{\bibfnamefont{A.}~\bibnamefont{Neronov}},
  \bibinfo{journal}{Astron. Astrophys.} \textbf{\bibinfo{volume}{525}},
  \bibinfo{pages}{A154} (\bibinfo{year}{2011}), \eprint{1011.1177}.

\bibitem[{\citenamefont{Kachelrie{\ss}
  et~al.}(2014)\citenamefont{Kachelrie{\ss}, Moskalenko, and
  Ostapchenko}}]{Kachelriess:2014mga}
\bibinfo{author}{\bibfnamefont{M.}~\bibnamefont{Kachelrie{\ss}}},
  \bibinfo{author}{\bibfnamefont{I.~V.} \bibnamefont{Moskalenko}},
  \bibnamefont{and} \bibinfo{author}{\bibfnamefont{S.~S.}
  \bibnamefont{Ostapchenko}}, \bibinfo{journal}{Astrophys. J.}
  \textbf{\bibinfo{volume}{789}}, \bibinfo{pages}{136} (\bibinfo{year}{2014}),
  \eprint{1406.0035}.

\bibitem[{\citenamefont{Kachelrie{\ss}
  et~al.}(2019)\citenamefont{Kachelrie{\ss}, Moskalenko, and
  Ostapchenko}}]{Kachelriess:2019}
\bibinfo{author}{\bibfnamefont{M.}~\bibnamefont{Kachelrie{\ss}}},
  \bibinfo{author}{\bibfnamefont{I.~V.} \bibnamefont{Moskalenko}},
  \bibnamefont{and} \bibinfo{author}{\bibfnamefont{S.~S.}
  \bibnamefont{Ostapchenko}} (\bibinfo{year}{2019}), \eprint{1904.05129}.

\bibitem[{\citenamefont{Ostapchenko}(2008)}]{Ostapchenko:2006nh}
\bibinfo{author}{\bibfnamefont{S.}~\bibnamefont{Ostapchenko}},
  \bibinfo{journal}{Phys. Rev.} \textbf{\bibinfo{volume}{D77}},
  \bibinfo{pages}{034009} (\bibinfo{year}{2008}), \eprint{hep-ph/0612175}.

\bibitem[{\citenamefont{Ostapchenko}(2011)}]{Ostapchenko:2010vb}
\bibinfo{author}{\bibfnamefont{S.}~\bibnamefont{Ostapchenko}},
  \bibinfo{journal}{Phys. Rev.} \textbf{\bibinfo{volume}{D83}},
  \bibinfo{pages}{014018} (\bibinfo{year}{2011}), \eprint{1010.1869}.

\end{thebibliography}

\end{document}